\documentclass{iau}
\usepackage{graphicx}

\title[Flux rope formation] 
{Observations of flux rope formation prior to coronal mass ejections}

\author[Lucie M. Green \& Bernhard Kliem]   
{Lucie M. Green$^1$
 \and Bernhard Kliem$^2$}

\affiliation{$^1$Mullard Space Science Laboratory, UCL, Holmbury St. Mary,
Dorking, Surrey, UK 
 \\[\affilskip]
$^2$University of Potsdam, Institute of Physics \& Astronomy,
Potsdam, Germany}

\pubyear{2013}
\volume{300}  
\jname{Nature of Prominences and Their Role in Space Weather}
\editors{B. Schmieder, J.-M. Malherbe, \& S. T. Wu, eds.}
\doi{10.1017/S1743921313010983}

\begin{document}

\maketitle

\begin{abstract}
Understanding the magnetic configuration of the source regions of 
coronal mass ejections (CMEs) is vital 
in order to determine the trigger and driver of these events.
Observations of four CME productive active regions are presented here, which 
indicate that the pre-eruption magnetic configuration is that of a magnetic flux rope.
The flux ropes are formed in the solar atmosphere
by the process known as flux cancellation and are stable for several hours 
before the eruption. The observations also 
indicate that the magnetic structure that erupts is not the entire
flux rope as initially formed, raising the question of
whether the flux rope is able to undergo a partial eruption
or whether it undergoes a transition in specific flux rope configuration
shortly before the CME.
 
\end{abstract}

\firstsection  

\section{Introduction}

Twisted bundles of magnetic field lines known as flux ropes are central to all
models of coronal mass ejections. They are invoked as being present either
before the eruption onset 
(e.g. \cite[T\"{o}r\"{o}k \& Kliem 2005]{Toeroekkliem05})
or being formed during the eruption itself
(e.g. \cite[Antiochos et al. 1999]{Antiochos_etal99}). 
In light of this, one way to
discriminate between these two sets of CME models is to 
determine the pre-eruption magnetic
configuration. However, there is an inherent difficulty in the
confident identification of flux ropes as it is not currently possible to
directly measure the magnetic field above the photosphere/chromosphere.
Instead, proxies for the presence of flux ropes need to be used, either from
 solar observations or from reconstructions of the coronal magnetic field 
using the photospheric field
as the boundary condition.

One well developed observational approach to investigate the pre-eruption
magnetic configuration is to study so-called sigmoidal
active regions (\cite[Rust \& Kumar 1996]{Rustkumar1996}). 
Sigmoidal regions contain S shaped 
EUV and X-ray emission structures and are regions that have a very 
high likelihood 
of producing a coronal mass ejection. Sigmoids 
are seen as predictors
of an eruption (\cite[Canfield \etal\ 1999]{Canfieldetal99}).
The sigmoid can be S or reverse S shaped, 
depending on the chirality of the magnetic field in which it 
forms (\cite[Pevtsov \etal\ 1997]{Pevtsovetal97}). 
\cite[Rust \& Kumar (1996)]{Rustkumar1996} suggested a link 
between sigmoids and kinking flux ropes, which has been a lively area of 
research ever since. 
Theoretical expectations from modelled
flux ropes show that layers of enhanced current, and presumably heating,
are located at the interface between a flux rope and its surrounding 
magnetic arcade (\cite[Titov \& D{\'e}moulin 1999]{Titovdemoulin99}). 
These layers should appear S shaped when viewed from above, building the 
case that sigmoids represent flux ropes in the solar atmosphere.

Recent investigations into the magnetic configuration of sigmoids
have shown support that, at least a sub-set of sigmoids, do indeed
indicate the presence of a magnetic flux rope.
\cite[Green \& Kliem (2009)]{Greenkliem2009}  
showed that continuous S shaped threads that make an inverse crossing of the
photospheric
polarity inversion line (PIL) in their centre strongly suggest field lines 
that spiral around the flux rope. 
These threads additionally cross the PIL in the normal direction in
the sigmoid's elbows. They trace 
field lines at the periphery of the rope and are likely to extend down to
the lower atmosphere, where they form bald patches.

As well as forming in active regions, sigmoids have been observed 
on a smaller scale in X-ray bright points 
(\cite[Mandrini \etal\ 2005]{Mandrinietal05})
and on a larger scale in the quiet Sun 
(\cite[Jiang \etal\ 2007]{Jiangetal07}).
It has been suggested that there are different types of sigmoid
relating to their lifetime (transient and long-lived) or detailed
observational appearance (continuous S threads or double J's that overall
look S shaped) (\cite[Pevtsov 2002a]{Pevtsov2002a}).


In summary, flux ropes can be investigated in 
the solar atmosphere prior to a coronal mass ejection by using
observations of sigmoids. 
These features provide an opportunity to follow the evolution of the 
magnetic configuration and investigate how the flux rope forms, the magnetic 
flux content of the rope and the specific magnetic configuration.
In this paper we focus on 
flux rope formation in a small sample of sigmoidal active regions 
in the days leading up to a coronal mass ejection.

\section{Observations of flux rope formation}

Here we present the evolution of four active regions; 
(1) NOAA region 10930 that was observed on the Sun during December 2006, 
(2) NOAA region 8005 that was seen during December 1996, (3) an un-numbered region
that was seen in February 2007 and (4) NOAA active region 10977 that was on
the disk during December 2007. This sample includes one region 
which forms a sigmoid during a flux emergence event and three
regions which show sigmoidal structure forming during the
decay phase of the region. 

\subsection{NOAA active region 10930}
Active region 10930
rotated over the eastern limb of the Sun on 5 December 2006. The region
contained a large negative polarity sunspot with a corresponding dispersed 
and spotless positive magnetic field to the east. Immediately to the south of 
the negative polarity sunspot was a smaller magnetic bipole. 
On 10 December 2006, new flux began to emerge in the location of the 
smaller bipole. The positive polarity of the emerging flux underwent an
eastward motion and a strong counter 
clockwise rotation of up to 540 degrees between 10 and 14 December 
(\cite[Min \& Chae 2009]{Minchae2009}). The positive polarity of the emerging 
flux was directly next to the pre-existing negative spot meaning
that flux cancellation was likely to be taking place.

The XRT/Hinode (\cite[Golub \etal\ 2007]{Golubetal07}) data show that as 
soon as the flux emergence begins,
the magnetic field of the positive sunspot formed
connections with the pre-existing negative sunspot producing a magnetic arcade
between them. The shear in this arcade
field increased rapidly, most likely due to 
the rapid eastward motion of the emerging flux and 
its strong counter-clockwise rotation.
By 23:15 UT on 11 December the arcade field had taken on the appearance of 
a 'double J' configuration and by 19:00 UT on 12 December a
continuous S shaped structure was seen. 
See the top row of Fig. \ref{fig1} for the coronal evolution of
this region. On 
13 December 2006 a GOES X3.4 class flare occurred in the sigmoidal region
at around 02:15 UT. A coronal
mass ejection was associated with this flare. 


\subsection{NOAA active region 8005}
NOAA active region 8005 had a bipolar magnetic configuration. 
The region emerged on the far side of the Sun
and as it rotated over the limb it was already in its decay phase
with dispersed magnetic polarities and no sunspots. The photospheric
field exhibited dispersal of the magnetic polarities and no episodes
of flux emergence. On many occasions opposite polarity fragments 
approached the active region's polarity inversion line and cancelled.
See \cite[Green \& Kliem (2009)]{Greenkliem2009} 
and \cite[Mackay \etal\ (2011)]{Mackayetal11} for details of the
evolution of the photospheric magnetic field in this region. 
In the days leading up to the eruption the active region loops as seen in the soft
X-ray data had an overall S shape which evolved from a sheared arcade, to a
double J shape (by 19 December 07:24 UT) to a sigmoid with continuous S shaped 
threads being
observed by 10:37 UT on 19 December. See the second from top
 row of Fig. \ref{fig1} for the coronal evolution of
this region. The active region produced an
eruption and a GOES C2.3 class flare on 19 December 1996 at 15:21 UT. 

\subsection{February 2007 region}

In February 2007 a sigmoid was observed in a very dispersed bipolar 
active region that had no NOAA number assigned to it. The region produced a
coronal mass ejection on 12 February 2007 beginning around 07:00 UT. There was
no flare emission associated with this eruption. In the days leading up to the
eruption the active region evolved from a double J configuration to having
continuous S shaped loops by end of the day on 11 February 2007.
See the second row from the bottom of Fig. \ref{fig1} 
for the coronal evolution of this region. The evolution
of the photospheric magnetic field showed ongoing 
dispersal and episodes of flux
cancellation at the polarity inversion line.
For more details on the evolution of this active region see
\cite[Savcheva \etal\ (2012)]{Savchevaetal12}.


\subsection{NOAA active region 10977}

NOAA active region 10977 had a bipolar configuration and the whole
lifetime of the active region was observed from emergence to 
dispersal into the surrounding quiet sun. The emergence began on 
3 December 2007 and the flux concentrations began to disperse
on 4 December 2007. During the decay phase
of the active region the photospheric field was dominated by
ongoing dispersal, an elongation of the polarities in the 
north-south direction and cancellation of flux along the polarity 
inversion line. 

\begin{figure} 
\begin{center}
 \includegraphics[width=12.5cm]{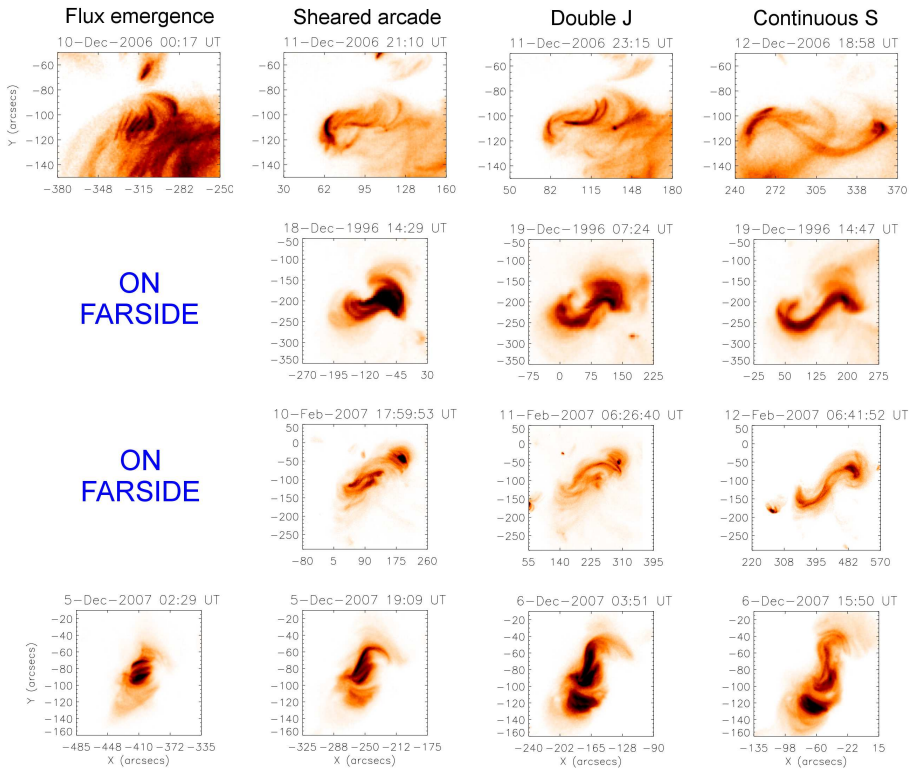} 
 \caption{Evolution of sigmoidal active regions showing the three phases of
 evolution from flux emergence, to sheared arcade, to 'double J' shaped loops
 and then finally the continuous S shaped threads of the sigmoid.
 Top row: NOAA region
 10930. Second row from the top: NOAA region 8005. Second row from the bottom:
 un-numbered region observed on the disk during 
 February 2007. Bottom row: NOAA region 10977.}
   \label{fig1}
\end{center}
\end{figure}

Hinode/XRT data show that the active region loops
appeared to have relatively little shear during the emergence phase
and into 5 December after the emergence had ceased. The shear began
to build up during the decay phase when the loops became much more
aligned to the polarity inversion line. Early on 6 December the
region appeared as a region of double J shaped loops with some remnant
arcade field in the south. Then, by 6 December 15:50 UT continuous 
S shaped sigmoidal threads were seen in the region. See the bottom 
row of Fig. \ref{fig1} for the coronal evolution of
this region. On 7 December at around 04:20 UT the region produced 
a coronal mass ejection. For a more detailed description of the evolution 
of this active region see \cite[Green \etal\ (2011)]{Greenetal2011}.

\section{Filament formation}

All of the above active regions exhibited the formation of a filament along 
the polarity inversion line where the sigmoid formed.
Big Bear Solar Observatory H-alpha data 
show that a filament was present in NOAA active region 10930 by 11 December 18:00 UT.
In NOAA region 8005 the filament had formed by 18 December 17:40 UT.
In the un-named region seen on the disk in February 2007, Kanzelh{\"o}he 
Observatory data show
that there was a filament present
by 10 February 2007 12:00 UT. In NOAA region 10977 Hinode/SOT H-alpha data show
that the filament was forming by 5 December around 20:00 UT. The filaments
in all these regions formed during the phase where the coronal arcade
was becoming more sheared. The presence of a filament indicates low lying 
loops that have dips, or twisted field lines which can support the dense plasma
against gravity.

\section{Coronal mass ejections from the sigmoidal regions}

\begin{figure} 
\begin{center}
 \includegraphics[width=9cm]{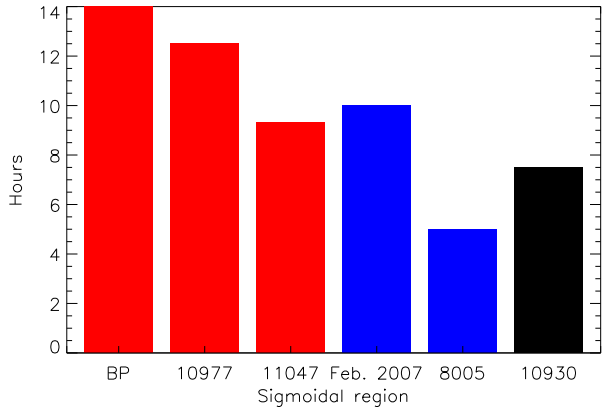} 
 \caption{Bar chart showing the time between first observation of continuous S
 shaped loops in the regions and the time of the coronal mass ejection. The colours
 indicate different region characteristics. Red: isolated bipolar regions that
 have been followed from emergence to eruption and which produce the sigmoid
 and the eruption
 in their decay phase. Blue: isolated bipolar regions that emerged on the
 far side of the Sun and which produced the sigmoid and eruption during their
 decay phase. Black: a multipolar active region tracked from a flux 
 emergence episode to the sigmoid formation and eruption and which formed the sigmoid
 during the flux emergence phase.}
   \label{fig2}
\end{center}
\end{figure}

The observation of the sigmoid allows the magnetic configuration to be probed
and a time for the formation of the flux rope to be identified. The S shaped field
lines can indicate the presence of magnetic field lines with dips which
suggests
the presence of a flux rope, even if it is not fully formed. 
The time between sigmoid (and flux rope) 
formation and the onset of
the coronal mass ejection is given in Fig. \ref{fig2}. To increase
the study size this figure 
also includes the sigmoid and coronal mass ejection that occurred 
in the bright point study of
\cite[Mandrini \etal\ (2005)]{Mandrinietal05} and the sigmoidal region 
NOAA 11047 that  produced a coronal mass ejection and that was studied in 
\cite[Savcheva \etal\ (2012)]{Savchevaetal12}.
The colours indicate whether the region was observed from the emergence
phase, and hence whether all activity is seen, or whether the region
emerged on the far-side so that aspects of the photospheric 
and coronal evolution may have been missed. In all cases the flux 
rope exhibits a stable phase of 
several hours between its formation and eruption.

In all four regions of this study the coronal mass ejection does not 
involve the filament material, as has been previously shown when sigmoidal 
active regions produce coronal mass ejections
(\cite[Pevtsov 2002b]{Pevtsov2002b}). The events presented here
show that the S shaped loops also do not rise during the eruption.
Instead, the erupting structure 
in three out of the four sigmoidal regions is a 
faint linear or loop-like feature that rises up as the flare loops
brighten underneath. See Fig. 3 in 
\cite[Su \etal\ (2007)]{Suetal07} for region 10930, 
\cite[McKenzie \& Canfield (2008)]{McKenziecanfield2008} for the case in
February 2007 and 
\cite[Green \etal\ (2011)]{Greenetal2011} for region 10977.

\section{Discussion}

There is growing observational support that a flux rope is present before 
the onset of a coronal mass ejection in some cases
(see \cite[Aulanier \etal\ 2010]{Aulanieretal10}, 
\cite[Green \etal\ 2011]{Greenetal2011} and references therein).
Active regions that show a sigmoidal structure have been fruitful
in revealing these flux ropes and their formation mechanism.
The observations presented here strongly suggest that flux ropes in the 
solar atmosphere are built by the flux cancellation mechanism proposed by 
\cite[van Ballegooijen \& Martens (1989)]{vBmartens1989}. 
Flux cancellation involves reconnection low down in the solar atmosphere 
between converging
opposite polarity fragments in a sheared arcade. The evolution 
of the coronal configuration is driven by the motions of the 
photospheric plasma and in this study is seen to pass through
three stages as arcade field evolves into
a flux rope. During the first stage shear in the coronal arcade field 
increases due to photospheric motions associated with flux emergence or flux 
dispersal and flux cancellation. Filaments start to form during this
stage. During stage two there is an accumulation of
a significant amount of axial
flux running along the polarity inversion line as flux
cancellation, further shearing and/or rotation of the magnetic polarities 
takes place. Remnant arcade field takes on the appearance of
two J's either side of this axial flux. In stage three, flux 
cancellation produces field
lines that are twisted around the axial flux and which contribute poloidal flux to
the rope. This flux cancellation scenario appears to be relevant 
in dispersed and isolated bipolar active regions in their decay phase,
where the flux rope forms along the polarity inversion 
line of the bipole, or during the flux emergence phase of a multipolar active
region where the flux rope forms along the polarity inversion line 
between neighbouring bipoles which are butted up against each other.
The observations of the evolutionary phases suggest that sigmoids 
do not exhibit different
types, rather their appearance evolves as the magnetic configuration changes. 

The details of the eruptions from these regions are also important for understanding
the pre-eruption magnetic configuration and the aspects of this 
configuration that are involved in the coronal mass ejection. 
The eruptions from these regions do not involve either the continuous S 
shaped threads or the filament. Since both are likely to involve field lines that
are located in, or extend down to, the dense plasma of the lower atmosphere, 
it is not surprising that they are immobile. The flux rope
cannot
erupt in its entirety in this situation. 
The observations show that the structure which does erupt is instead
a collection of loops that connect between 
the elbows of the S shaped threads and which can be seen before the 
onset of the eruption. These loops have been called 
a linear feature and tentatively been
associated with the core of the erupting flux rope 
in \cite[Green \etal\ (2011)]{Greenetal2011}.
However, in \cite[Aulanier \etal\ (2010)]{Aulanieretal10} the same feature
is interpreted as being a consequence of heating in a current shell
above the rope and called an erupting loop-like feature.
Such an erupting structure was also seen by \cite[Moore \etal\ (2001)]{Mooreetal01}
but has become
more frequently observed with Hinode/XRT due to its large dynamic range and
also with SDO due to the increased plasma temperature coverage. See, for example,
\cite[Liu \etal\ (2010)]{Liuetal10}
and \cite[Zharkov \etal\ (2011)]{Zharkovetal11}. 

The observations presented here suggest that the flux rope is 
able to partially erupt, allowing the accumulated axial flux to 
escape and become the erupting linear feature,
whilst leaving the flux associated to the S shaped threads and filament 
behind. In other cases
the flux rope may evolve
prior to the eruption so that its underside is detached from the 
lower atmosphere, allowing it to erupt fully, as demonstrated in the simulations of 
\cite[Aulanier \etal\ (2010)]{Aulanieretal10}.

\end{document}